# PHOTOVOLTAIC POWER GENERATION IN THE STELLAR ENVIRONMENTS

T.E.Girish   and   S.Aranya,
Department of Physics,University College,
Thiruvananthapuram  695 034
INDIA

E mail :  tegirish5@yahoo.co.in

## ABSTRACT

In  this paper we have studied the problem of photovoltaic power generation near selected stars in the solar neighborhood. The nature of the optical radiation from a star will depend on its luminosity,HR classification and spectral characteristics. The solar cell operation in the habitable zones of the stars is similar to AM1.0 operation near earth.The current space solar cell technology can be adopted for power generation near G,K and M type stars. Silicon solar cells with good near IR response are particularly suitable in the environments of M type stars which are most abundant in the universe. . Photovoltaic power generation near binary stars  like *Sirius* and *Alpha Centauri* is also discussed.

                  M type stars-Alpha Centauri

# 1. Introduction

There is an increasing demand for the interstellar space missions since several planetary systems are recently discovered around other stars[1]. Even though there are many studies discussing about power sources for interstellar travel [2] the problem of photovoltaic power generation using stellar radiations is not explicitly investigated so far. In this paper we have studied photovoltaic power generation near selected stars in the solar neighborhood.

Solar cell operation in the stellar environments depends on several factors such as:-

  (i)   The optical radiation power distribution near the star as a function of its luminosity.
  (ii)  The spectral characteristics of stellar optical radiations and finding suitable solar cells.
  (iii) The radiation environment ( high energy particles and radiation) near the stars and its possible impact on solar cell operation.
  (iv)  Solar cell operation in a close binary star environment.

We have attempted to address the above problems in this study making use of relevant astronomical observations and the current solar cell technology.

## 2. Optical radiation power distribution near the stars as a function of stellar luminosity

The optical radiation power distribution near the stars is mainly dependent on the luminosity. Let $\alpha$ be the ratio of the luminosity of the given star ($L_s$) to that of our Sun ($L_o$).

$$\alpha = L_s/L_o \qquad (1)$$

The power of stellar radiation ($P_s$) will change with radial distance from the star as

$$P_s = \alpha S/r^2 \qquad (2)$$

Here S is the solar constant measured above earth's atmosphere ( 1370 W/m^2)

In Table 1 we have given α values for some selected nearby stars [3] along with their distances from the sun [4]. Except for *Sirius B* ( a white dwarf) all of them are main sequence stars in the HR classification.

Within our solar system the power of the solar radiation between 1 to 20 A.U ( the orbit of Pluto) decreases from 1370 W/m^2 to 34 W/m^2. Between 1-20 A.U from a faint star like the *Barnard's star* ( α = 0.0004) the Ps varies from 5.48 W/m^2 to 13.7 mW/m^2. For a bright star like *Sirius A* ( α = 23) the same varies between 31510 to 78.8 W/m^2.

The optimum distance (d) for photovoltaic power generation near a star is that distance where the power of the radiation from the star equals the solar constant (S).
This optimum distance can be calculated from the simple relation [ 5 ]

d ( A.U ) =   SQRT ( α )            ( 3)

For the stars given in Table 1 we have calculated d values and included in the same table.

The minimum power required for photovoltaic power generation for LILT solar cells is 0.01 W/m^2 [6-7] .The distance at which the stellar radiation power decreases to the above value can be considered as the distance limit from a star for photovoltaic power generation according to the current solar cell technology. This is also calculated for the stars in Table 1.

### 3. Spectral characteristics of stellar optical radiations and photovoltaic power generation

For main sequence stars with surface temperatures between 3000 to 30,000 K blackbody radiation laws are applicable to the stellar spectral irradiance[ 8 ].Using the Wein's displacement law the wavelength of maximum spectral irradiance($\lambda$m) from a star can be calculated [9] from its surface temperature T.

$\lambda$mT=   0.29 X 10 E-3.         (4)

Using (4) we have calculated $\lambda$m values for the nearby stars and shown in Table 2 along with the surface temperatures of these stars [10].

The band gap of the solar cell materials will decide the spectral response of the solar cell.concerned. The spectral irradiance of the star should match with the spectral response of the solar cell illuminated by the stellar radiations.The good near IR response of the silicon solar cells will match the spectral irradiance characteristics of M type stars whose $\lambda m$ lies in that region. Polycrystalline silicon solar cells [ 11 ] are suitable for M stars whose $\lambda m$ is between 900-1000 nm and single crystal Si solar cells is suitable for stars with $\lambda m$ between 800-850 nm [12].

For photovoltaic power generation near G or K type stars with surface temperatures between 5000-6000 K, Group III-V solar cells like Ga As with good spectral response around 500 nm will be a good selection [13].

For stars with surface temperatures > 6500 K, $\lambda m$ lies either in the blue or UV radiation of the spectrum. Near these stars we need to employ spectral down converters along with suitable solar cells for photovoltaic power generation [ 14-15]. This technology is however only an emerging one .

### 4. Photovoltaic power generation near binary stars

It is interesting to study the photovoltaic power generation in the close binary star environments . The technological problem is equivalent to a solar cell operation when simultaneously illuminated by two independent light sources with different intensity and spectral irradiance. *Sirius* and *Alpha Centauri* are our nearest visual binary systems. These double stars actually move around the centre of mass of the system. However the apparent orbit of the less massive star can considered as an ellipse around the heavier companion [16]. The distance between *Sirius A* ( heavy) and *Sirius B* varies between 11 AU to 35 AU.The distance between *Alpha centauri A* ( heavy) and *Alpha centauri B* varies between 8.5 AU to 31.5 AU [10].

An observer located along the line joining the stars in a binary system will receive maximum radiation from both the stars. In Table 3 we have given the calculated values of intensity(power) of radiations received by such an observer separately for the Sirius and the Alpha centauri binary systems. The distance given in the first column of this table is measured from less massive star ( Alpha centauri B or Sirius B) concerned. Here 0.05 AU and 0.68 AU are the optimum distances for photovoltaic power generation respectively for Sirius B and Alpha centauri B . In the third and fifth columns of this table, quantities given within parentheses correspond with occasions of maximum separation from these binary stars.

## 5. Discussion

Nearest stars ( within 12 light years from sun) are selected in this study which are likely to be destinations of future interstellar space missions. Majority of these are main sequence stars which are relatively more abundant and whose physical properties are better undertood.Further blackbody radiation laws can be applied to the spectral irradiance from these stars.

Luminosity of the stars and their spectral irradiance characteristics of its radiations are basic factors which decide photovoltaic power generation near them. Distances for optimum photovoltaic power generation (defined section 2) is identical to the location of habitable zones of these stars. At these distances solar cell operation resembles AM1.0 operation near earth. The maximum distance limits for photovoltaic power generation based on the current LILT solar cell technology is found to vary between 1-1000 AU from nearby stars.

The relative abundance of stars belonging to the different spectral classes according to astronomical observations are: M type ( 80 % ) , K type ( 8 % ) ,G type ( 3.5 % ) ,F type ( 2 % ) ,A type (0.7 % ) ,B type ( 0.1 % ) and O type ( 0.0001 % ) [ 17].

It is interesting to note that silicon the most abundant solar cell material in earth is found to be suitable for photovoltaic power generation near M type stars which are observed to be the most abundant in our universe. Further due to good near IR response of Si solar cells their efficiency will be likely to be higher when operated with M type stellar radiations compared to radiations from sun like stars.

The stellar radiation environment is mainly controlled by phenomena such as galactic cosmic rays,stellar winds and manifestations of stellar activity changes such as flares. a The radiation environments of stars like *Alpha Centauri* [18] are either similar or less hazardous compared to near earth space environments. However radiation environments of bright stars like *Sirius A* is likely to be mare hazardous compared to near earth space [19] . The knowledge gathered from radiation properties of space solar cells operated within our solar system will be helpful in this context [20].

The relative luminosities of stars in a binary system and their varying distances of separation are important in deciding photovoltaic power generation in their environments, For occasions of minimum separation between *Alpha cenrtauri A* and *B* it is found that intensity of radiation from the former star becomes significant only for distances > 2.5 AU from the later star.

During occasions of maximum separation of these stars the contribution from *Alpha centauri A* near its companion is negligible. In contrast with this due to the very high luminosity of *Sirius A* its radiation contribution near its companion
*Sirius B* is always significant as evident from Table 3.

With the current space technology the time required for interstellar travel even to the nearest stellar neighbor *Alpha centauri* is at least 20 years. Laboratory models of stellar photovoltaic generation with artificial light sources resembling stellar radiations [9] can provide us more knowledge on this topic. The details of such a study will be addressed in a future publication.

**6.Conclusions**

- (i) The distance limits to photovoltaic power generation in the stellar environments is decided by the luminosity of the star concerned. The photovoltaic power generation in the habitable zones of stars is similar to AM1.0 solar cell operation near earth.

- (ii) The current space solar cell technology can be adopted for power generation near G,K and M type stars. Silicon crystalline solar cells with good near IR spectral response is particularly suitable for operation near M type stars.

- (iii) The relative luminosities of stars in a binary system and their varying distances of separations are important in controlling the photovoltaic power generation near these stars.

## References

1] S.Marchi,*Astrophys.J.* 666 ( 2007 ) 475-485.

[2]  Robert.L.Forward, *J.Spacecraft* 22 ( 1985) 345

[3]  www.stellar_database.com

[4]  www.cosmobrain.com

[5]  M.Hart, *Icarus*  37 ( 1979) 825

[6]  T.E.Girish, *Sol.Energy Mat. Sol.Cell.* 90 ( 2006) 825-831.

[7]  T.K.Kerslake,D.A.Sheiman,in: Proceedings of the Third international energy conversion conference,TM-2005-213988 ( NASA Glenn Res. Centre,USA,2005) 1-47.

[8]  Martin Johnson ,Astronomy of stellar energy and decay,(Dover pub,Newyork,1959)

[9]  G.A.Rechtseiner and J.A.Ganski, *The Chem. Edu*. 3 ( 1998) S1430.4171(98)04230.7

[10]   www.en.wikipedia.org/wiki

[11]   W.G.J.H.M.van Sark,A.Meijenuk,R.E.I.Schropp,J.A.M.van Roosmalen,E.H.Lysen, *Sol.Energy Mat.Sol.Cell. ,* 87 ( 2005) 395-409.

[12] .F.Strobl,P.G.Uebele,K.H.Tentscher,R.Kern,K.D.Rasch,K.P.Bogus,T.Robben, G.Lo Roche,in: Poceedings of the 28$^{th}$  Photovoltaic Specialists Conference, ( IEEE,Anchorage,2000) 1289-1292.

[13]  L.L.Kazmerki,*Renew.Sust.Energy Rev*. 1 ( 1997) 71-170.

[14]  B.S.Richards, *Sol.Energy Mat.Sol.Cell.*  90 ( 2006) 1189-1207.

[15]  T.Trupke,M.A.Green,P.Wurfel, *J.Appl.Phys*. 92 ( 2002) 1668-1674.

[16]  Michael Zeilek,Astronomy-The evoloving universe,Fifth Edition ( John Wiley& Sons, Newyork,1988).

[17]   www.atlasoftheuniverse.com

[18]   B.E.Wood,J.L.Linsky,H.R.Muller,G.P.Zank,Astrphys. J. 547 ( 2001) L 49-L52.

[19]  P.Bertin,H.J.G.L.M. Lamers,A.Vidal-Madjar,A.Ferlet,R.Lalkmen, Astron.Atrophys. 302  ( 1995) 899.
[20] A.Suzuki,M.Kaneiwa,T.Saga ,S.Matsuda *IEEE Trans.Electr. Dev*. 46 (1999)

**Table 1. Optimimum and maximum distances for photovoltaic power generation for selected nearby stars as a function of stellar luminosity**

| Name of star | Distance from sun (AU) | Luminosity relative to sun (α) | Optimum Distance (AU) | Maximum Distance (AU) |
|---|---|---|---|---|
| 1. Alpha cent A | 4.35 | 1.567 | 1.25 | 413.82 |
| 2. Alpha cent B | 4.35 | 0.46 | 0.68 | 251 |
| 3. Barnard's star | 5.98 | 0.0004 | 0.02 | 7.4 |
| 4. Wolf 359 | 7.78 | 0.00002 | 0.0045 | 1.66 |
| 5. Lalande 21185 | 8.26 | 0.00568 | 0.075 | 27.94 |
| 6. Sirius A | 8.55 | 25 | 4.8 | 1775 |
| 7. Sirius B | 8.55 | 0.0025 | 0.05 | 18.51 |
| 8 Epsilon Eridani | 10.7 | 0.3 | 0.548 | 202.73 |
| 9. Procyon A | 11.38 | 7.73 | 2.78 | 1029 |

**Table 2. Spectral characteristics of nearby stars and solar cells suggested for photovoltaic power generation using corresponding radiations**

| Name of star | Spectral class | Surface temperature(K) | Wavelength of maximum irradiance ( nm) | Solar cells suggested |
|---|---|---|---|---|
| 1.Alpha cent A | G2 | 5800 | 500 | Gr. III-V cells |
| 2.Alpha cent B | K4 | 5300 | 547 | Gr. III-V cells |
| 3. Barnard's star | M4 | 3134 | 925 | Polycryst. Si |
| 4.Wolf 359 | M6 | 3500 | 828 | Monocryst Si |
| 5. Lalande 21185 | M2 | 3400 | 852 | Monocryst Si |
| 6.Sirius A | A1 | 9940 | 292 | Down converters & solar cells |
| 7. Sirius B | White dwarf | 25200 | 115 | Down converters & solar cells |
| 8.Epsilon Eridiani | K2 | 5100 | 568 | Gr. III-V cells |
| 9.Procyon A | F5 | 6530 | 444 | Down converters & solar cells |

**Table 3. Maximum optical power due to binary star companions along the line joining the stars in the binary systems of *Sirius* and *Alpha Centaturi* at times of minimum and maximum separations.**

| Distance from Sirius B or Alpha Cent B ( AU) | Optical power from Sirius A ( W/m2) | Optical power From Sirius B ( W/m2) | Optical power from α Cent A ( W/m2) | Optical power from α cent B ( W/m2) |
|---|---|---|---|---|
| 0.05 | 441.3 ( 31.85) | 1370 | | |
| 0.68 | | | 20.16 ( 1.82) | 1370 |
| 1 | 560.2 ( 33.87) | 2.43 | 21.47 ( 1.86) | 630.2 |
| 2 | 745.8 ( 36.21) | 0.86 | 26.5 ( 1.97) | 157.6 |
| 3 | 1042 ( 38.79) | 0.38 | 33.5 ( 2.1) | 70.02 |
| 4 | 1556 ( 41.67) | 0.21 | 43.8 ( 2.2) | 39.38 |
| 5 | 2572 ( 44.87) | 0.14 | 59.6 ( 2.4) | 25.21 |
| 6 | 5041.6(48.5) | 0.095 | 85.9 (2.6) | 17.5 |
| 7 | 14004 ( 52.5) | 0.070 | 134.2 ( 2.7) | 12.86 |